\documentclass[prb,twocolumn,showpacs]{revtex4} 
\usepackage{graphicx}
\usepackage{amsmath,amssymb,bm} 
\usepackage{epstopdf}
\usepackage{booktabs}
\usepackage{xspace,color,soul,hyperref}
\newcommand{\euform}{EuCl\ensuremath{_3\cdot}6H\ensuremath{_2}O\xspace}

\newcommand{\eutrans}{{\ensuremath{^7}F\ensuremath{_0\rightarrow ^5}D\ensuremath{_0}}\xspace}
\newcommand{\iso}[1]{\ensuremath{{^{#1}}}} 
\newcommand{\tplus}{\ensuremath{^{3+}}\xspace}

\newcommand{\ctwo}{C\ensuremath{_{2}}\xspace}
\newcommand{\grounds}{\ensuremath{^7}F\ensuremath{_0}\xspace}
\newcommand{\excits}{\ensuremath{^5}D\ensuremath{_0}\xspace}

\newcommand{\AAA}{{\AA}\xspace}

\newcommand{\degree}{\ensuremath{^\circ}\xspace}
\newcommand{\mb}[1]{\ensuremath{\mathbf{#1}}}
\newcommand{\mbh}[1]{\ensuremath{\mathbf{\hat{#1}}}}

\begin{document}

\title{A method for assigning satellite lines to crystallographic sites in rare earth crystals}

 \author{R. L. Ahlefeldt}
\affiliation{Centre for Quantum Computation and Communication Technology, Research School of Physics and Engineering, The Australian National University, Canberra, ACT 0200, Australia}
\author{W. D. Hutchison}
 \affiliation{School of Physical, Environmental and Mathematical Sciences, The University of New South Wales, Canberra, ACT 2600, Australia}
 \author{N. B. Manson} \author{M. J. Sellars}
 \affiliation{Centre for Quantum Computation and Communication Technology, Research School of Physics and Engineering, The Australian National University, Canberra, ACT 0200, Australia}

\date{\today}
\begin{abstract}
We describe an experimental technique for associating the satellite lines in a rare earth optical spectrum caused by a defect with the rare earth ions in crystal sites around that defect. This method involves measuring the hyperfine splitting caused by a magnetic dipole-dipole interaction between host ions and a magnetic defect. The method was applied to Ce\tplus:\euform to assign 13 of the outermost 22 satellite lines to sites. The assignments show that the optical shift of a satellite line is loosely dependent on the distance to the dopant. The interaction between host and dopant ions is purely dipole-dipole at distances greater than 7~\AA, with an additional contribution, likely superexchange, at distances less than 7~\AA. 
\end{abstract}


\maketitle
\section{Introduction}
The optical spectra of rare earth ions in crystals commonly display satellite lines: multiple weak lines arrayed on either side of the optical main line and separated from it by up to terahertz. These satellite lines arise from rare earth ions in sites neighboring to point defects such as chemical impurities. Frequently, the defects causing satellite line structure in a rare earth crystal are other rare earth impurities, and in the case where the optically active rare earth ion itself is a dopant, satellite lines arise from neighboring pairs of dopant ions. 

Satellite lines provide an opportunity to probe the direct neighborhood of a defect, and are often used to identify, and characterize, defects in crystals \cite{cone93, jaaniso94, yamaguchi99}. In rare earth doped crystals, ion-pair satellite lines allow energy-transfer and resonant electronic interactions between closely separated ions to be studied \cite{vial79}. Because each pair site is due to ions with a fixed separation, the distance dependence of interactions can be determined from measurements on many different satellite lines. In Pr\tplus:LaF$_3$, for example, energy transfer between Pr\tplus ions separated by more than 4.2\AA\ is predominantly due to an electric dipole dipole interaction \cite{vial82, hegarty82}, but at shorter distances, there is a substantial additional contribution from another mechanism, attributed to superexchange \cite{vial82, morgan85}. A better understanding of these nearest-neighbor interactions is useful for practical applications, such as lasers, because even if the proportion of rare earth ions in pair sites is small, they can be the dominant source of up-conversion and fluorescence quenching processes \cite{birgeneau69}.

Satellite lines have also been suggested as a way of making a frequency-addressed, ensemble-based quantum computing system that can be scaled to moderate numbers of qubits \cite{sellars04}. This quantum computing scheme requires a crystal stoichiometric in the rare earth ion of interest, such as Eu\tplus, and doped lightly with another rare earth. Satellite lines in such a material arise from Eu\tplus ions surrounding dopant ions, with each satellite line  due to an ensemble of Eu\tplus ions in a unique position relative to the dopant ions. Using these satellite lines as qubits has the twin advantages of a high density of ions in each ensemble qubit and very strong, homogeneous interactions between qubits, which arise because the separation of ions in one qubit from their partners in another qubit are of the order of angstroms. Additionally, rare earth ions in solids can have very long optical and nuclear coherence times, and have long been considered a good platform for quantum computing \cite{pryde00, lukin00, ichimura01,shahriar02, ohlsson02}.

A common difficulty when studying interactions between rare earth ions in satellite lines, which also hampers experimental demonstrations of quantum computing in stoichiometric materials where these interactions are used to enact multi-qubit gates, is the difficulty in determining the crystallographic site around a dopant to which a particular satellite line is due. Site assignments have only been achieved in a small number of high symmetry materials, in which it is possible to exploit the symmetry to assign sites \cite{jaaniso94, wietfeldt85,jouart85}. In this paper, we describe a method for  assigning satellite lines caused by a magnetic defect to crystallographic sites that is applicable to all symmetries: satellite lines are assigned by measuring the effect of the magnetic dipole-dipole interaction between defect and rare earth ions on the ground state hyperfine structure of the rare earth ion.

To demonstrate the method, site assignments were performed for satellite lines caused by Ce\tplus in \euform. As the satellite line structure of rare earth doped \euform changes only by a scaling factor for different dopants, these assignments apply to all rare earth doped \euform crystals. \euform was chosen because it is a good candidate for the stoichiometric quantum computing scheme described above, as it can have long coherence times when fully deuterated and has the narrowest optical inhomogeneous linewidth of any stoichiometric solid, allowing a high density of ions in each prepared ensemble qubit. The site assignments presented here will allow the interactions between different satellite lines to be measured and characterized, and the performance of multi-qubit gates enacted using these interactions to be estimated.

\section{Theoretical background}
\euform is a monoclinic crystal with $P2/n$ crystal symmetry and \ctwo symmetry at the Eu\tplus site \cite{bel`skii65, kepert83}. The two Eu\tplus isotopes, \iso{151}Eu and \iso{153}Eu, have nuclear spin $I = \frac{5}{2}$ and the singlet optical ground (\grounds)  and excited (\excits) states are split into three doubly degenerate hyperfine levels in zero magnetic field, with splittings of the order of 50~MHz \cite{ahlefeldt13a}. The hyperfine structure of a single optical transition can be described by a spin Hamiltonian of the form \cite{macfarlane87}:
\begin{equation} \label{eqn:fithamil}
 H= \textbf{B}\cdot \textbf{M}
\cdot\hat{\textbf{I}}+\hat{\textbf{I}} \cdot \textbf{Q} \cdot \hat{\textbf{I}}
\end{equation}
with $\textbf{B}$ the magnetic field, $\hat{\textbf{I}}$ the nuclear spin operator,
\begin{align}
 M & =R(\alpha, \beta, \gamma_m) \left[\begin{array}{lll}
g_x & 0 & 0\\
0 & g_y & 0\\
0 & 0 & g_z
      \end{array}\right]R^T(\alpha, \beta, \gamma_m)\label{eqn:zeeman}\\
\intertext{ the enhanced nuclear Zeeman tensor, and}
Q&= R(\alpha, \beta, \gamma_q)\left[\begin{array}{lll}
-E & 0 & 0\\
0 & E & 0\\
0 & 0 & D
\end{array}
\right] R^T(\alpha, \beta, \gamma_q) \label{eqn:quadrupole}
\end{align}
the effective quadrupole tensor. In Equations \eqref{eqn:zeeman} and \eqref{eqn:quadrupole}, $R(\alpha, \beta, \gamma)$ is a rotation matrix in three Euler angles, of the form given in Ref. \onlinecite{longdell06}. The \ctwo symmetry of the Eu\tplus site constrains one of the three principal axes of both $\textbf{M}$ and $\textbf{Q}$ to lie along the crystal \ctwo axis. The orientation of the \ctwo axis relative to the $z$ axis of the coordinate system used is given by the Euler angles $\alpha$ and $\beta$. The spin Hamiltonian of the optical ground and excited states of Eu\tplus in \euform has been characterized previously.

\begin{figure}
\centering
\includegraphics[width = 0.35\textwidth]{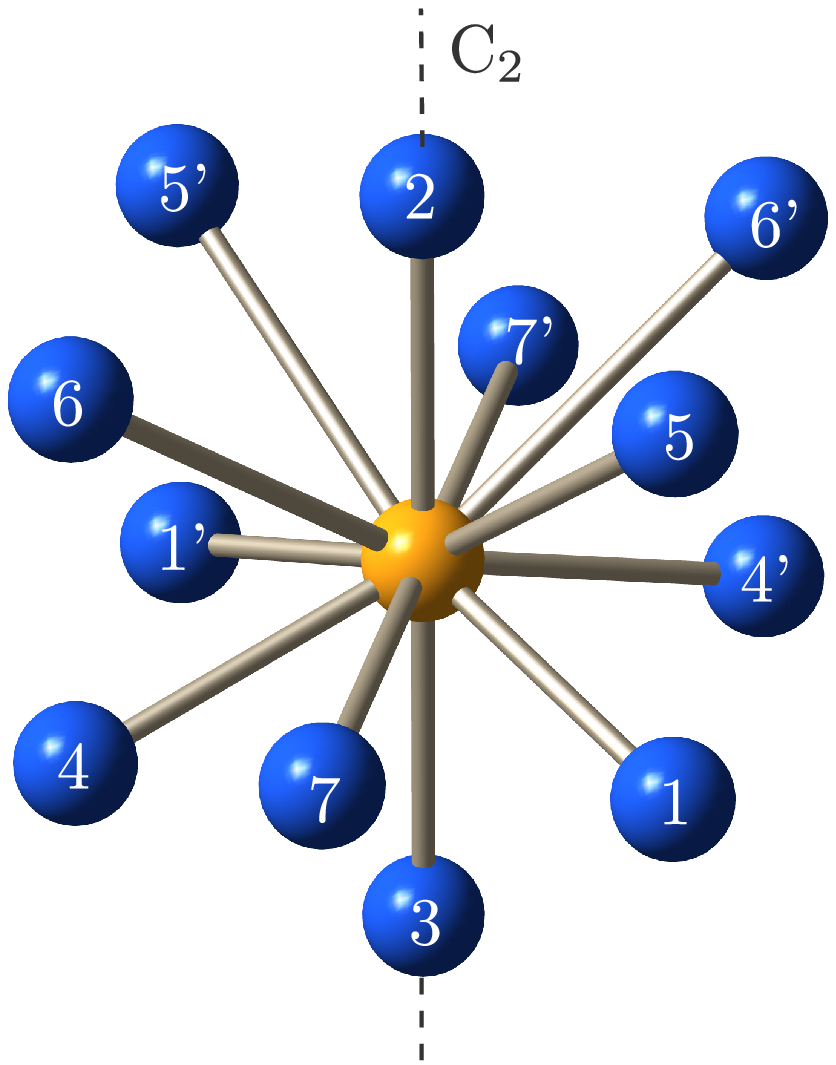}
\caption{ \label{fig:eupos} The first shell of Eu\tplus ion positions around a central dopant ion in \euform, with ion sites labeled according to their distance from the dopant. The crystal \ctwo axis runs through sites 2 and 3, and the dashes indicate equivalent sites related by a 180\degree rotation about the \ctwo. The ions are separated from the dopant by between 6.4 \AA\ (site 1) and 7.9 \AA (site 7). The closest ions in the next shell are 9.7 \AA\ from the dopant.}
\end{figure}

To assign satellite lines to crystallographic sites in \euform, the crystal was doped with Ce\tplus, a Kramers dopant. The presence of a Kramers dopant has been shown to lead to superhyperfine splitting of the ground state hyperfine levels that differs for different satellite lines, with the size of the splitting commensurate with a magnetic dipole-dipole interaction between the electronic moment of the Kramers dopant ion and the nuclear moment of the Eu\tplus ions \cite{ahlefeldt10}. Assuming a magnetic dipole-dipole interaction,  the spin Hamiltonian of the coupled Ce\tplus--Eu\tplus system is the spin Hamiltonian in Equation \eqref{eqn:fithamil} with two additional terms, the Zeeman Hamiltonian of the dopant and the dipole dipole interaction:
\begin{equation}
H^i= \mathbf{B}\cdot \mb{M}\cdot\mbh{I} +   \mbh{I}\cdot\mb{Q}\cdot\mbh{I}   +   \hat{\mb{B}} \cdot \mb{M_{Ce}} \cdot \mbh{S}   + H_{dd}^i \label{eqn:fithamilkr}\,.
\end{equation}
In this equation $\mbh{S}$ and $\mb{M_{Ce}}$ are the electron spin operator and Zeeman tensor of Ce\tplus respectively and $H_{dd}^i$ is the magnetic dipole-dipole interaction. The dipole-dipole interaction is dependent on the position of the Eu\tplus ion relative to the dopant, given by the site index $i$, meaning that different satellite lines, which are due to different crystallographic sites, will have different hyperfine splittings. The Eu\tplus ion positions around a dopant ion in \euform are shown in Figure \ref{fig:eupos}. The sites are labeled according to their distance from the dopant ion.

 The dipole-dipole interaction can be written
\begin{align}
 H_{dd}^i = \frac{\mu_0h}{4\pi|\bm{r}_i|^3}&\left[(\mb{M}\cdot\mbh{I}) \cdot(\mb{M_{Ce}}\cdot\mbh{S})\right. \nonumber \\
&\left.-3(\bm{\hat{r}_i}\cdot\mb{M}\cdot\mbh{I})(\bm{\hat{r}_i}\cdot\mb{M_{Ce}}\cdot\mbh{S})\right]\,, \label{eqn:fitdd}
\end{align}
where $\bm{r_i}$ is the position vector joining the Eu\tplus and Ce\tplus sites, which can be determined for each site from the crystal structure \cite{kepert83}. The Zeeman tensor $\mb{M_{Ce}}$ of Ce\tplus, which has the same form as Equation \eqref{eqn:zeeman}, has been measured in YCl$_3\cdot$6H$_2$O\cite{schulz67}. The Zeeman tensor components were measured to be  $g_x = 37.4$~GHz/T, $g_y = 10.9$~GHz/T, and $g_z = 32.2$~GHz/T, with one principal axis ($g_z$) lying along the crystal C$_2$ axis, while the others lie in the (010) plane with the axis corresponding to $g_y$ 21\degree clockwise from [100]. As YCl$_3\cdot$6H$_2$O is isostructural with \euform and the two rare earth ions are fairly similar in radius, the electronic magnetic moment of Ce\tplus in \euform will be very similar to that of Ce\tplus in YCl$_3\cdot$6H$_2$O.

\section{Experimental method}
Satellite lines were assigned to crystallographic sites in 0.1\% Ce\tplus:\euform by recording the hyperfine splitting of each satellite line as an external magnetic field was rotated about the sample and fitting this rotation pattern to the spin Hamiltonian of Equation \eqref{eqn:fithamilkr} to determine the site index $i$. 

The experimental method used was similar to that reported previously\cite{ahlefeldt13a}. The 0.1\%Ce\tplus:\euform crystal used was grown from a water solution using 99.999\% CeCl$_3\cdot$7H$_2$O and \euform starting materials. The crystal was mounted in a set of three-axis superconducting coils with the [100] axis approximately parallel to  the $x$ axis of the superconducting coils and the laser direction, and the [010] direction approximately aligned with the $-z$ axis of the coils. The laser was polarized along the crystal \ctwo axis, [010]. 

The sample was cooled to approximately 2~K in a helium bath cryostat, and the external magnetic field was rotated in a spiral in 200 steps about the sample given by
\begin{equation}
 \mathbf{B} = \left[\begin{array}{c}
                     -B_0\sqrt{1-t^2}\sin{6\pi t}\\
		     B_0 t\\
		     -B_0\sqrt{1-t^2}\cos{6\pi t}		     
                    \end{array}\right]
\label{eqn:spiral}
\end{equation}
for $t$ ranging from $-1$ to $+1$.
The hyperfine structure was recorded with Raman heterodyne double resonance spectroscopy \cite{mlynek83}. A Coherent 699-29 dye laser tuned to the \eutrans transition at 517148.5~GHz was used to supply the optical field needed for Raman heterodyne spectroscopy, while a small 8-turn coil mounted around the crystal inside and coaxial with the $x$ axis of the superconducting coils was used to supply an rf field. An rf signal of approximately 0.1~mT was generated with an HP Spectrum Analyzer connected to a 40~W rf amplifier. The combination of optical and rf fields results in coherent emission at the sum and difference frequencies of the two fields, which was detected as a beat on the transmitted laser light by the same spectrum analyzer that was used to generate the rf field.

\begin{figure}
\centering
\includegraphics{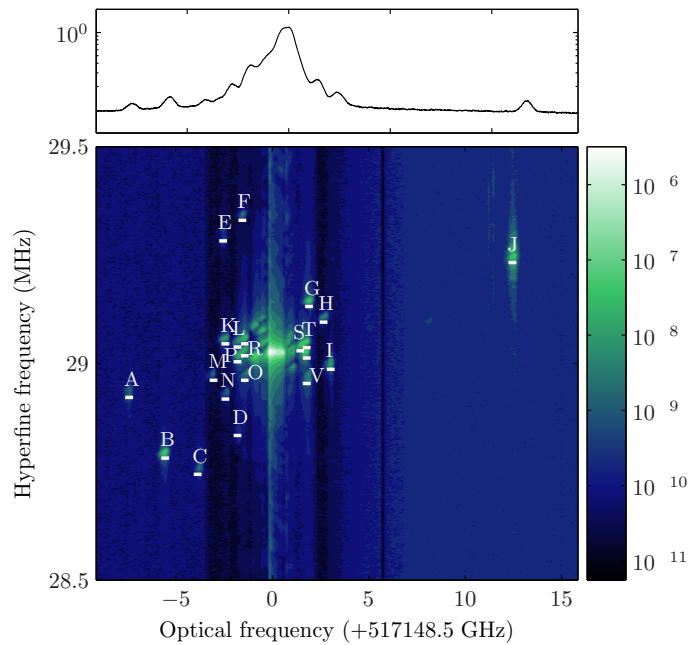}
\caption{ \label{fig:ceeudr} Excitation spectrum of  0.1\%Ce\tplus:\euform with a logarithmic vertical axis (top) and double resonance spectrum of the 29 MHz ground state transition of \iso{151}Eu\tplus in 0.1\%Ce\tplus:\euform (bottom). The labeled satellite lines are those rotation patterns were recorded for. The color axis in this, and all subsequent double resonance spectra, is logarithmic.}
\end{figure}
A two-dimensional Raman heterodyne double resonance spectrum of 0.1\% Ce\tplus:\euform is shown in Figure \ref{fig:ceeudr}. The satellite lines for which rotation patterns were recorded are labeled in this figure. Initially, rotation patterns for the 27 and 29~MHz hyperfine ground state transitions of \iso{151}Eu\tplus were recorded with $B_0 = 21$~mT in the center of the optical line at 517148.5~GHz, where the signal is due to Eu\tplus ions unperturbed by Ce\tplus. This allowed the orientation of the magnet coordinate frame relative to that of the previous spin Hamiltonian characterization \cite{ahlefeldt13a} to be determined. Following this, rotation patterns were recorded for only the 29~MHz hyperfine ground state transition for the 22 labeled satellite lines in Figure \ref{fig:ceeudr} with $B_0 = 10.5$~mT. A smaller applied magnetic field was used for the satellite lines because it results in larger differences between the rotation patterns of different satellite lines.

\section{Results}
\begin{figure}
\centering
\includegraphics{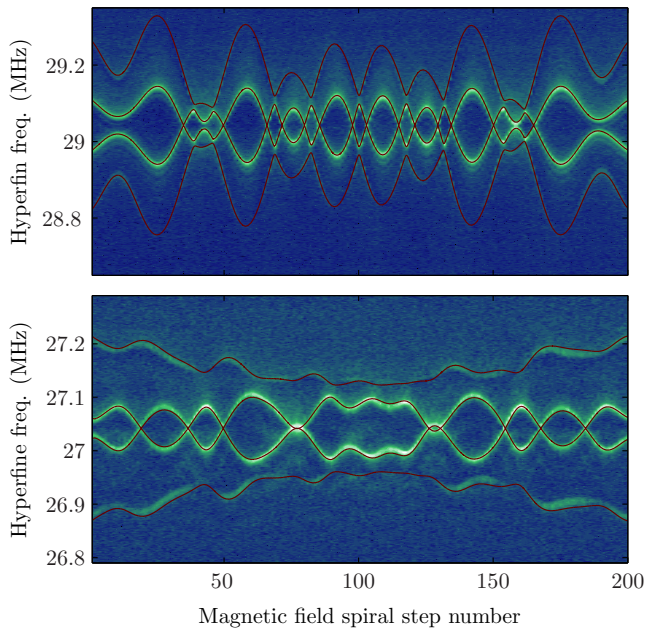}
\caption{ \label{fig:cemainrot} Experimental rotation pattern for the main line of 0.1\%Ce\tplus:\euform about the ground state transition of \iso{151}Eu\tplus. The red lines are a fit to the spectrum.}
\end{figure}
Raman-heterodyne-detected rotation patterns about the 27 and 29 MHz ground state transitions of \iso{151}Eu\tplus in 0.1\% Ce\tplus:\euform are shown in Figure \ref{fig:cemainrot}. These rotation patterns are described by Equation \eqref{eqn:fithamil} and the spin Hamiltonian parameters given in Ref. \onlinecite{ahlefeldt13a}. The red lines in this figure are a fit to the data using this spin Hamiltonian. The parameters in the fit are three Euler angles $\alpha_B$, $\beta_B$, and $\gamma_B$ that describe a rotation matrix $R(\alpha_B, \beta_B, \gamma_B)$ that rotates the magnetic field $\mathbf{B}$ into a coordinate frame fixed with respect to the crystal axes: ([100],$[100]\times[010]$, $[0\bar{1}0]$). The fit shown in Figure \ref{fig:cemainrot} gave $\alpha_B = 5.00\degree$, $\beta_B = 7.79\degree$ and $\gamma_B = -6.51\degree$, which is reasonable given that aligning the crystal with the magnetic field coils by eye is only accurate to $\pm10\degree$. While this fit gives the alignment of the magnetic field relative to the crystal axes, there is still an uncertainty of $\delta = \pm10\degree$ in the orientation of the $\mathbf{M}$ and $\mathbf{Q}$ tensors in the (010) plane. This cannot be determined in measurements of the spin Hamiltonian of the main line, but does effect rotation patterns of satellite lines. The theoretical fits shown in the remainder of this paper use $\delta =-8\degree$, as this gives the best fit to most satellite lines.

\begin{figure}
\centering
\includegraphics{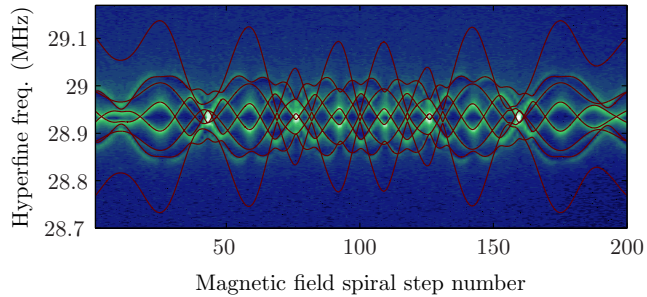}
\caption{ \label{fig:cerota} Rotation pattern of 0.1\% Ce\tplus:\euform at 517140.95~THz (line A). Red lines show the best fit theoretical pattern: site 2 or 3, the two sites on either side of the dopant along the C$_2$ axis. While these sites are crystallographically distinct, they have the same rotation patterns. }
\end{figure}
\begin{figure}
\centering
\includegraphics{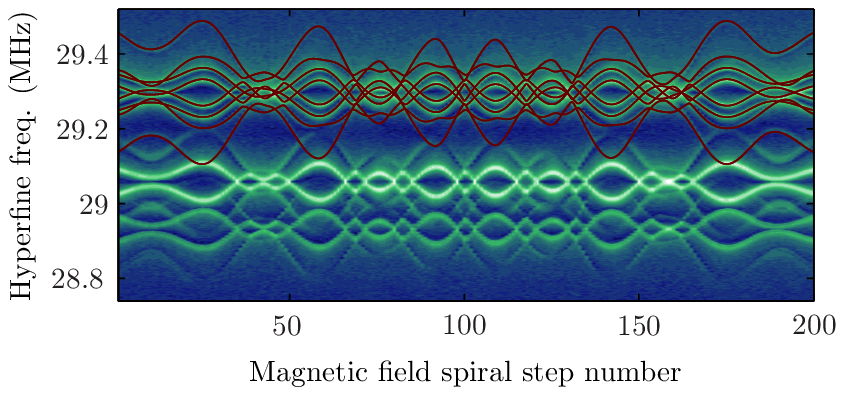}
\caption{ \label{fig:cerote} Rotation pattern of 0.1\% Ce\tplus:\euform at 517145.79~THz (lines E, K and N). Line E is assigned to site 7}
\end{figure}
\begin{figure}
\centering
\includegraphics{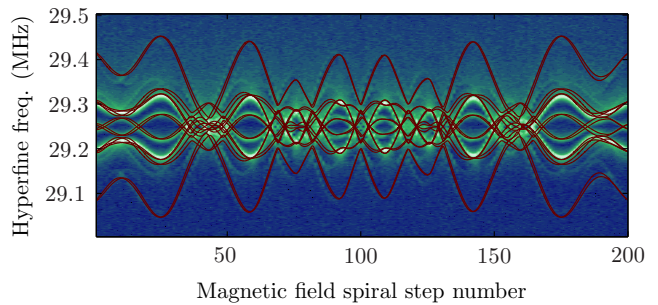}
\caption{ \label{fig:cerotj} Rotation pattern of 0.1\% Ce\tplus:\euform at 51760.67~THz (line J). This is the outermost satellite line, and corresponds to the closest ion site, site 1. The reasons for the poor fit to this satellite line are discussed in the text.}
\end{figure}

Three example satellite line rotation patterns are shown in Figures \ref{fig:cerota}--\ref{fig:cerote}  The experimental  patterns differ substantially from that of the main line, Figure \ref{fig:cemainrot}, with the 4 lines in that pattern split into either 8 or 16 lines by the dipole-dipole interaction. Eight lines arise because the interaction is sensitive to the orientation of the Ce\tplus magnetic moment, while some sites show 16 lines because the interaction differs slightly for the two \ctwo symmetric positions contributing to that satellite line. 

Also shown in Figures \ref{fig:cerota}--\ref{fig:cerote} are the theoretical pattern for the crystallographic site that gives the best match to the experimental rotation pattern. As the theoretical pattern has only one unknown parameter, the site index $i$, the fits were determined by comparing the theoretical and experimental patterns by eye. Satellite line A is due to one of the two sites on the \ctwo axis, sites 2 and 3. Although these two sites are crystallographically distinct, they have the same rotation pattern. Satellite line E is due to site 7, while line J is due to site 1. Two other satellite lines, lines N and K, are visible in Figure \ref{fig:cerote}, but these have patterns very similar to the main line and cannot be assigned to a crystallographic site.  This does not necessarily imply that these lines are due to sites a long way away from the dopant: the dipole-dipole interaction is strongly dependent on the orientation of the site and there are a number of sites at distances of less than 15 \AA\ that show almost no dipole-dipole splitting in the rotation pattern. Even some satellite lines that do show a dipole-dipole splitting can be difficult to assign to a site as many of the outer sites (separated by more that 10 \AA) have very similar patterns to each other. The site assignments that were made are summarized in Table \ref{tab:assign}.
\begin{table}
\begin{ruledtabular}
\begin{tabular}{lll}
Line & Site  & Distance from dopant (\AA) \\ \hline
A &  2 or 3 & 6.52\\
B &  9 or 10 & 10.27\\
C  &  2 or 3 & 6.52\\
D &  4 & 6.74\\
E &  7 & 7.94\\
F &  9 or 10 & 10.27 \\
G &  15 &  12.10\\
H &  8 & 9.66\\
I &  11 & 10.43\\
J &  1 & 6.36\\
L &  19 or 22 & 12.89 (19), 13.27 (22) \\
M &  16 20 or 21 & 12.63 (16), 13.06 (20, 21) \\
V &  5 & 7.57\\
\end{tabular}
\caption{\label{tab:assign} Association between ion sites and satellite lines determined from experimental rotation patterns. Only those sites that could be assigned are listed. The sites are labeled by their distance from the dopant. The actual site positions can be generated from the crystal structure \cite{kepert83}, and the positions of the nearest seven sites to the dopant ions are shown in Figure \ref{fig:eupos}.}
\end{ruledtabular}
\end{table}

\section{Discussion}
In the model used to fit the hyperfine structure of satellite lines, the magnetic interaction between host and dopant ions was assumed to be dipole-dipole. For the outer satellite lines (sites 7 and above), such as line E (Figure \ref{fig:cerote}), the dipole-dipole interaction fit the data well, justifying this assumption. However, for lines J (Figure \ref{fig:cerotj}) and line D, the fit to the data was poor, although sufficient to definitively assign these lines to sites 1 and 4, respectively.  Lines A (Figure \ref{fig:cerota}) and C, which were assigned to the two sites 2 and 3 on the dopant's \ctwo axis, also do not fit as well as most other lines. The four anomalous satellite lines A,C, D and J arise from the four innermost sites, with separations to the dopant of between 6.36 and 6.74 \AA. The next closest sites (which give rise to lines V and E) are $\approx 7.8$ \AAA away from the dopant, and show no anomalous behavior. This suggests that a short range interaction is contributing to the nearest sites. There are two possible sources of this interaction: crystal strain and exchange with the dopant. These are described below.

The strain caused by the Ce\tplus dopant can affect the rotation pattern in two different ways. Firstly, it could modify the quadrupole and Zeeman tensors of nearby Eu\tplus ions. The quadrupole tensor is certainly modified to some extent, because the satellite lines have different zero field hyperfine frequencies to the main line, but the difference is fairly small, $< 0.5 \%$. Any substantial modification of the Zeeman tensor can be ruled out by looking at the rotation pattern of Pr\tplus:\euform. Pr\tplus has a similar radius to Ce\tplus, but no electronic magnetic moment in the ground state because the orbital angular momentum is quenched, so if the dopant was modifying the quadrupole and Zeeman tensors, it would be expected that the rotation patterns on the innermost ion sites in Pr\tplus:\euform would be different from the main line pattern. We have measured rotation patterns for lines A and J in Pr\tplus:\euform; both these lines have patterns identical to the main line.

The second way the dopant-induced strain can affect the rotation pattern is by shifting the positions of the surrounding Eu\tplus ions from their unperturbed positions, thus altering the magnetic dipole-dipole interaction between the dopant and the Eu\tplus ion. The shift of nearest neighbor ions can be estimated by looking at the rare earth separations in the isomorphic RECl$_3$.6H$_2$O (RE = Nd, Gd, Lu) \cite{habenschuss80,habenschuss80a,habenschuss80b}. The rare earth separations in these materials differ by less than 1\% from that in \euform, suggesting that a Ce\tplus dopant is unlikely to shift the position of neighboring Eu\tplus ions by more than 1\%. This is much smaller than the 10\% shifts required to explain the discrepancy between experimental and theoretical rotation patterns.

The final option is an exchange interaction between the Ce\tplus ion and the nearby Eu\tplus ions, most likely superexchange, which has been seen between rare earth ions with separations as large as 10~\AA \cite{birgeneau68}. The sharp cutoff in the interaction at $\approx 7$~\AAA does provide some evidence that the interaction is superexchange, as this is a feature of this interaction.

The site assignments presented here for \euform show how the optical and hyperfine frequencies of a satellite line are related to the spatial position of the ions in that line. In general, the satellite lines in \euform that are shifted the most in optical or hyperfine frequency are the closest sites to the dopant, but some satellite lines break this trend: line B has the third largest optical shift, but is due to a site more than 10 \AA\ away from the dopant, while the fact that the line corresponding to site 6 (one of the first shell sites) is missing suggests that its optical shift is very small. This demonstrates that, while the interaction between dopant and host ions that results in optical and hyperfine shifts is distance-dependent, the distance dependence is not strong. More concrete conclusions about the interaction and the distortion field of the dopant could be made by modelling the optical and hyperfine satellite structure using the site assignments.

While the site assignment method has been demonstrated for a stoichiometric Eu\tplus crystal doped with another rare earth, it is more generally applicable. It can be used to assign satellite lines caused by any defect with a large magnetic moment. The method is useful for both low and high symmetry materials, and can assign most of the satellite lines in any material to a single crystallographic site. In all materials with non-centrosymmetric rare earth sites, there will be a small number of lines that can only be assigned to one of two sites related by inversion symmetry, as the spin Hamiltonian itself has inversion symmetry. This is shown here for \euform: lines A and C are due to sites 2 and 3, but because the two sites are related by inversion symmetry and so have identical rotation patterns, it cannot be determined which line corresponds to which site. In centrosymmetric crystals, sites related by inversion no longer have distinct optical transition frequencies, and all sites can be assigned. 

In the rare earth doped crystals more commonly studied in rare earth spectroscopy site assignments could be achieved by co-doping the crystal with a Kramers dopant with a similar radius. For instance, a Pr\tplus doped crystal could be co-doped with Ce\tplus. This would lead to an additional set of satellite lines due to Ce\tplus--Pr\tplus pairs, which can be correlated with the Pr\tplus--Pr\tplus lines because the optical shift of a satellite line is dependent on ion radius \cite{fricke79}. The rotation pattern technique could be performed on the additional satellite lines to assign them to sites. 

The criterion that allows a satellite line to be used as a frequency addressed qubit is that it is optically well resolved from other qubit satellite lines. Where two lines occur at the same optical frequency, one can be used. Of the 13 satellite lines whose spatial positions were determined in this paper, at least 11 could readily be used as qubits. As these satellite lines are due to the inner shells of ions around the Ce\tplus dopant, the distances between the 11 qubit lines are small, ranging from 6.4 to 26.3 \AA, and interactions between the qubits can be expected to be strong. The next step in this work is to measure these interaction strengths between different satellite lines.

\section{Conclusion}
Assigning satellite lines caused by a magnetic defect in a rare earth crystal to the specific rare earth ion positions around that defect is possible by utilizing the magnetic dipole-dipole interaction between the Eu\tplus nuclear spin and the defect electronic spin. In \euform, this method was used to assign 13 of the 22 outermost satellite lines to crystallographic sites. These assignments showed that most of the outer satellite lines are due to the innermost site positions around the dopant. While the interaction between dopant and host ions separated by more than 7~\AA\ is wholly magnetic dipole-dipole, at distances of less than 7\AA, there is a small contribution to the interaction from another mechanism, likely superexchange.

\begin{acknowledgments}
This research was conducted by the Australian Research Council Centre of Excellence for Quantum Computation and Communication Technology (Project number CE110001027).
\end{acknowledgments}


%

\end{document}